\documentclass[twocolumn]{aastex631}

\begin{document}

\title{Anomalous emission from Li- and Na-like ions in the Corona heated via Alfv\'{e}n wave}

\author[0000-0002-1043-9944]{Takuma Matsumoto}
\affiliation{Centre for Integrated Data Science, Institute for Space-Earth Environmental Research, Nagoya University, Furocho, Chikusa-ku, Nagoya, Aichi 464-8601, Japan}
\affiliation{National Astronomical Observatory of Japan, 2-21-1 Osawa, Mitaka, Tokyo 181-8588, Japan}








\begin{abstract}

The solar ultraviolet intensities of spectral lines originating from Li- and Na-like ions have been observed to surpass the expectations derived from plasmas with coronal approximation. 
The violation of the coronal approximation can be partially attributed to non-equilibrium ionization (NEI) due to dynamic processes occurring in the vicinity of the transition region.
To investigate the impact of these dynamics in Alfv\'{e}n-wave-heated coronal loop, a set of equations governing NEI for multiple ion species was solved numerically in conjunction with 1.5-dimensional magnetohydrodynamic equations. 
Following the injection of Alfv\'{e}n waves from the photosphere, the system undergoes a time evolution characterized by phases of evaporation, condensation, and quasi-steady states.
During the evaporation phase, the ionization fractions of Li- and Na-like ions were observed to increase when compared to the fractions in ionization equilibrium, which lead to the intensity enhancement of up to 1.6. 
This over-fractionation of Li- and Na-like ions was found to be induced by the evaporation process.
While collisions between shocks and the transition region temporarily led to deviations from ionization equilibrium, on average over time, these deviations were negligible.
Conversely, under-fractions of the ionization fraction led to intensity reduction of down to 0.9 during the condensation phase and the quasi-steady state.
Given the dependency of the over/under-fractionation on mass circulations between the chromosphere and the corona, these observations will serve as valuable benchmarks to validate not only Alfv\'{e}n wave models but also other existing mechanisms on coronal heating.
\end{abstract}



\section{Introduction} \label{sec:intro}

The investigation of extreme ultraviolet (EUV) emission lines stands as a foundational approach for comprehending high-temperature, low-density plasmas present in the solar corona. This comprehension serves as vital information essential to understanding coronal heating mechanisms. A substantial portion of deductions drawn from EUV line analysis relies on ionization equilibrium calculations. These calculations have been effectively applied to establish models for the solar corona and transition region, estimate chemical abundances, and propose diagnostic methodologies for assessing coronal density and temperature (see a review by \cite{2018LRSP...15....5D} and references therein).

Among the generally successful EUV observations, a noteworthy exception lies in the observations of Li- and Na-like ions, which exhibit abnormal behavior compared to the other ions, as well documented in \cite{2002A&A...385..968D}. In essence, the anomalous behavior refers to a consistent underestimation or overestimation, by substantial factors, of the theoretical intensities of lines emitted by these ions, following a Differential Emission Measure (DEM) analysis using lines from various isoelectronic sequences. This inconsistency has been noted by multiple researchers \citep{1971RSPTA.270...81B, 1972ApJ...178..527D}, and efforts to address it by incorporating density dependence in dielectric recombination and ionization from meta-stable levels initially appeared to resolve the anomalies \citep{1979ApJ...228L..89V,1981ApJ...245.1141R}. However, improvements in atomic databases and photometric accuracy revealed that these anomalies persist \citep{1995ApJ...455L..85J}. Despite the persistence of anomalies, notable improvements have been achieved in reducing the inconsistency by considering the effects of high densities, solar radiation, and charge transfer on ion formation \citep{2023MNRAS.521.4696D}.

The anomalous behavior exhibited by Li- and Na-like ions has prompted certain researchers to explore solutions beyond the assumptions commonly applied to numerous other spectral lines. One of the main assumptions is that all the ions are in a state of ionization equilibrium. However, when the plasma encounters a rapid increase in temperature, the assumption of instantaneous response in ion populations becomes invalid. This circumstance could lead to the potential existence of ions with relatively low charge at much higher temperatures than typically expected, consequently resulting in anomalous emission. Studies have investigated line formation during sudden temperature fluctuations, examining factors such as velocity redistribution, steady flow \citep{1989ApJ...338.1131N, 1990ApJ...362..370S, 2004ApJ...606.1239P}, and abrupt heating or cooling \citep{1993ApJ...402..741H, 2003A&A...401..699B, 2004A&A...425..287B}. Efforts with multi-dimensional simulations have been undertaken to reconcile the disparities observed between theoretical predictions and actual observations \citep{2013ApJ...767...43O, 2015ApJ...802....5O, 2016ApJ...817...46M}. However, performing multi-dimensional simulations remains a challenging task due to the thin transition region, where the majority of Li- and Na-like ions form.

The anomalous emission from Li- and Na-like ions, at least in part, appears to be linked to heating mechanisms and subsequent mass movements. Investigating this behavior holds potential importance in addressing the coronal heating problem. Reproducing this anomalous emission is important as a benchmark test for existing coronal heating models. This paper aims to explore the ability to reproduce the anomalous line formation process in a coronal model heated by Alfv\'{e}n waves. To accomplish this, we employed a 1.5-dimensional magnetohydrodynamics (MHD) model \citep{2004ApJ...601L.107M, 2004ESASP.575...80M} to predict the emissions from ions in non-equilibrium ionization (NEI). The hot corona is maintained via shocks produced through nonlinear mode conversions from Alfv\'{e}n waves \citep{1982SoPh...75...35H, 1999ApJ...514..493K}, and its behaviors in quasi-steady states \citep{2010ApJ...712..494A} and during long-term evolution \citep{2019ApJ...885..164W} have been extensively investigated. By solving a set of equations for NEI, we can quantitatively explore the behaviors of Li- and Na-like ions in the coronal loop with complex dynamics. Because the transition region, where most Li- and Na-like ions form, is a very thin layer, even small mass flows associated with evaporation, condensation, and shock propagation can occur in time scales shorter than the ionization and recombination time scale. This phenomenon may contribute to deviations in ionization fractions from ionization equilibrium, although detailed numerical simulations are required to quantitatively estimate the degree of departure in ionization fraction and the associated emissions.
Please note that our focus lies on the effects of bulk flows on NEI, although, in realistic situation, we can not decouple the impacts of alterations in charge states under high densities \citep{1979ApJ...228L..89V} and/or non-Maxwellian electrons \citep{2014ApJ...780L..12D, 2017SoPh..292..100D}. Including these effects will be the future studies.

This paper is organized as follows: Section 2 presents models and assumptions, Section 3 details our simulations and analysis with discussions, and Section 4 provides the conclusions of the study.

\section{Models and Assumptions} \label{sec:model}

In this study, we investigated the impact of NEI on emergent intensity by employing a coronal loop model heated by Alfvén waves. We concurrently solved the dynamic heating process governed by MHD equations and the evolution of the ion fraction. Subsequently, we reconstructed UV radiations based on the obtained ion fractions, electron density, and temperature.

We employed a 1.5-dimensional numerical model for the coronal loop, building upon the pioneering works of \cite{2004ApJ...601L.107M} and \cite{2004ESASP.575...80M}. This model naturally reproduces warm loops as a consequence of Alfv\'{e}n wave injection from the photosphere. The hot corona is achieved through MHD shocks generated by nonlinear mode conversion from Alfv\'{e}n waves \citep{1982SoPh...75...35H, 1999ApJ...514..493K}. 
The model's properties have been extensively investigated, including parameter dependencies \citep{2010ApJ...712..494A, 2010ApJ...710.1857M}, and differences from the nanoflare model \citep{2008ApJ...688..669A}.

The fundamental equations solved are as follows:
The equation of mass conservation:
\begin{equation}
\frac{\partial \rho A}{\partial t} + \frac{\partial \rho v_{\rm s} A}{\partial s} = 0, \label{eq_eom}
\end{equation}
the equation of momentum conservation along the field line:
\begin{eqnarray}
    \frac{\partial \rho v_s A}{\partial t} + \frac{\partial }{\partial s} \left[ \left( \rho v_s^2 + P_{\rm g} + \frac{B_\phi^2}{2} \right) A\right] \nonumber \\ 
    = \left( P_{\rm g} + \frac{\rho v_\phi^2}{2}\right) \frac{d A}{d s} - \rho g_{\rm s} A, \label{eq_eoms}
\end{eqnarray}
the equation of angular momentum conservation:
\begin{equation}
\frac{\partial \rho v_\phi A^{3/2}}{\partial t} + \frac{\partial}{\partial s} \left[ \left( \rho v_\phi v_{\rm s} - B_\phi B_{\rm s}\right) A^{3/2} \right] = \rho L_{\rm trq} A,\label{eq_eomp}
\end{equation}
the induction equation:
\begin{equation}
\frac{\partial B_\phi A^{1/2}}{\partial t} + \frac{\partial}{\partial s} \left[ \left( B_\phi v_{\rm s} - B_{\rm s} v_{\phi} \right) A^{1/2} \right] = 0, \label{eq_ie}
\end{equation}
and the equation of total energy conservation:
\begin{eqnarray}
\frac{\partial {\cal E} A}{\partial t} + \frac{\partial}{\partial s}  \left[ \left\{ \left( {\cal E} + P_{\rm g} + \frac{B_\phi^2}{2} \right) v_s - B_\phi B_{\rm s} v_\phi \right\} A \right] \nonumber \\
= - L_{\rm rad} A + Q_{\rm cnd} A - \rho v_{\rm s} g_{\rm s} A + \rho v_\phi L_{\rm trq} A^{1/2}. \label{eq_eot}
\end{eqnarray}
In these equations, $\rho$ and $P_{\rm g}$ represent mass density and gas pressure, respectively; $v_{\rm s}$ is the velocity along the field line; $v_\phi$ is the toroidal velocity; $B_\phi$ is the toroidal magnetic field strength normalized by $\sqrt{4\pi}$. ${\cal E}$ is the total energy density given by
\begin{equation}
    {\cal E} = \frac{1}{2} \rho \left( v_{\rm s}^2 + v_\phi^2 \right) + \frac{P_{\rm g}}{\gamma -1} + \frac{B_\phi^2}{2},
\end{equation}
where $\gamma$ is the ratio of specific heats and is taken to be $5/3$.

The variables $g_{\rm s}$ and $A$ represent gravitational acceleration along the field line and the cross-section of the flux tube, respectively. These variables depend on the shape of the flux tube, and we adopted the same shape as used in \cite{2004ApJ...601L.107M}. 
The radius of the flux tube, $r\equiv \sqrt{A/\pi}$, is described by the following set of equations:
\begin{eqnarray}
    r &=& \int \cos \theta ds, \\
    z &=& \int \sin \theta ds,
\end{eqnarray}
where $z$ denotes the length along the axis of the flux tube and $\theta$ is defined as
\begin{eqnarray}
    \theta = \theta_{\rm t} + (\theta_{\rm r} - \theta_{\rm t}) f_{\rm n}, \\
    \theta_{\rm r} = - \arctan \left( -\frac{4H_0}{r} \right), \\
    \theta_{\rm t} = \arctan \left[ k \cosh \left(\frac{z}{11 z_{\rm d}} \right) \right], \\
    f_{\rm n} = - \frac{1}{2} \left[ \tanh \left( \frac{z-z_{\rm d}}{H_0} \right) - 1 \right].
\end{eqnarray}
The parameters, $k, z_{\rm d}$, and $H_0$, are set to 3.48, 6$H_0$, and 200 km, respectively.
The field line below the chromosphere is significantly inclined from the vertical direction, leading to p-mode leakage.

Equation (\ref{eq_eot}) includes source terms accounting for radiation, thermal conduction, gravity, and torque. For radiation, we employed a composite model that considers both optically thick and thin radiative loss mechanisms \citep{2021A&A...656A.111S}. The radiative loss functions from optically thin plasma were estimated via CHIANTI \citep{2021ApJ...909...38D}, assuming photospheric element abundances. 
This choice was made to ensure consistency, as these abundances are also utilized in the EUV synthesis, as elaborated upon in subsequent paragraphs.
We did not account for the feedback of NEI in the radiative loss function, although it has been suggested that the impact of NEI on the radiative loss function is projected to be at most 5\% in dense atmospheres after solar flares \citep{1984SoPh...90..357M} or at most a factor of 2 to 4 in coronal loops \citep{1982ApJ...255..783M, 1993ApJ...402..741H}.

We selected the Spitzer-type thermal conduction. We did not use any flux limiter \citep[e.g.][]{2006SoPh..234...41K} because the heat flux in our simulations consistently smaller than the saturated flux. To reduce numerical diffusion near the lower transition region, we implemented a broadening technique for temperatures below $0.15$ MK, denoted as $T_{\rm c}$ \citep{2009ApJ...690..902L}. This technique involves adjustments to both the thermal conduction coefficient and radiative cooling. 
By implementing this technique, the width of the transition region below $T<T_{\rm c}$ will be broadened by a factor of $\sim (T_{\rm c}/T)^{5/2}$.
Consequently, we achieved a reduction in the relative temperature difference between adjacent grid points, $\Delta \ln T$, to less than 5\% for the current grid size, thereby ensuring the allocation of more than 20 grid points for the transition region. We will discuss the limitations of these modifications in subsequent sections.

The amplitude of the torque at the footpoints is denoted as $L_{\rm trq}$. This torque was enforced only at the footpoints, and its amplitude was determined such that the root mean square of the toroidal velocity amplitude reached $\sim$ 1 km s$^{-1}$. This velocity amplitude is consistent with the horizontal velocities in the granular cells \citep{1998ApJ...509..435V, 2010ApJ...716L..19M, 2012ApJ...752...48C, 2017ApJ...849....7O}.

We adopted an approximated equation of state in our model to include the effect of a change in molecular weight \citep{2014MNRAS.440..971M}. Although this slightly modified the atmospheric structure below the chromosphere ($T<10^4$ K) compared to constant molecular weight, the effects on the transition region and the corona should be subtle.

We investigated the impact of NEI for the elements C, N, O, Ne, Mg, Si, S, and Fe, as these elements harbor the most abundant ion species in the transition region and the corona. In this investigation, we present results specifically for C, N, O, Si, and S, which demonstrate notable deviations from ionization equilibrium.
In this regard, we solved a series of NEI equations for these elements, expressed as follows:
\begin{eqnarray}
  \frac{\partial N_{\rm i}}{\partial t} + \nabla \cdot \left(N_{\rm i} \mathbf{v} \right) \nonumber \\
  = N_{\rm e} \left( S_{\rm i-1} N_{\rm i-1} + \alpha_{\rm i+1} N_{\rm i+1} - \left( S_{\rm i} + \alpha_{\rm i} \right) N_{\rm i} \label{eq_nei} \right), 
\end{eqnarray}
where $N_{\rm i}$ denotes the number density of a specific element in the ith ionization stage while $N_{\rm e}$ indicates the electron number density. The coefficients on the right-hand side, $S_{\rm i}$ and $\alpha_{\rm i}$, indicate temperature-dependent collisional ionization and recombination rate coefficients obtained from CHIANTI atomic database 10.0.2. These rates are obtained in low density limit, although the density dependence removes some discrepancies between theory and observations on Li- and Na-like ions \citep{1979ApJ...228L..89V,1981ApJ...245.1141R,2023MNRAS.521.4696D}. Note that we did not include the feedback effects on MHD variables through the radiative cooling function.

We developed an original MHD code capable of conducting precise and stable simulations, even in the low beta region surrounding the transition region. To calculate numerical flux, we employed the HLL-approximated Riemann solver \citep{1991JCoPh..92..273E}. Furthermore, conservative variables were reconstructed in each cell using a 3rd order weighted essentially non-oscillatory (WENO) scheme and subsequently integrated in time through the 3rd order Arbitrary Derivative Riemann Problem (ADER) scheme \citep{2009JCoPh.228.2480B}. Considering that the time scale of thermal conduction is generally much shorter than that of dynamics, we adopted an operator split method, implicitly solving thermal conduction using the super time-stepping method \citep{2012MNRAS.422.2102M}. Because the shortest time scale in equations (\ref{eq_nei}) is often smaller than dynamical time scales, we implicitly solved the right-hand side of equations (\ref{eq_nei}).

We applied the same boundary and initial conditions as outlined in \cite{2010ApJ...712..494A}. In brief, the initial conditions maintained hydrostatic equilibrium up to a height of 2 Mm, above which an artificially denser atmosphere was assumed to avoid severe CFL condition. The numerical domain spanned 103 Mm, including 2 Mm of subphotospheric layers at both ends to mitigate artificial oscillations arising from the boundaries. Grid spacing started at 10 km for the initial 16 Mm from both boundaries and gradually increased to 100 km in the central region. 
We initially computed the entire evolution of coronal loops using this coarse grid spacing. 
We have confirmed that this coarse run converged well with 10 km grid size (see appendix).
Subsequently, we concentrated on three specific 20-minute time intervals and conducted additional calculations with finer grid sizes, reaching down to 1.25 km near the transition regions. This approach enabled us to reduce computational costs while effectively resolving the narrow transition region.
We also have conducted a convergence test for this fine grid simulation (see appendix).

\section{Results and Discussions}
In this study, we performed 1.5-dimensional MHD simulations of a coronal loop model subjected to Alfv\'{e}n wave heating. Simultaneously, we solved the NEI equations for selected elements. Subsequently, by tracking the dynamic evolution of temperature, electron density, and ionization fractions, we calculated the emergent intensity employing the CHIANTI database. Our results unveil that specific Li- and Na-like ions display higher intensities during ionizing process when contrasted with the predictions based on the coronal approximation: 
the plasma is optically thin, the plasma have Maxwellian distribution function, the plasma is in ionization equilibrium, ionization and recombination rate coefficients do not change with density (zero density limit in this study). 

\subsection{Properties of coronal model}

\begin{figure}
	\includegraphics[width=\columnwidth]{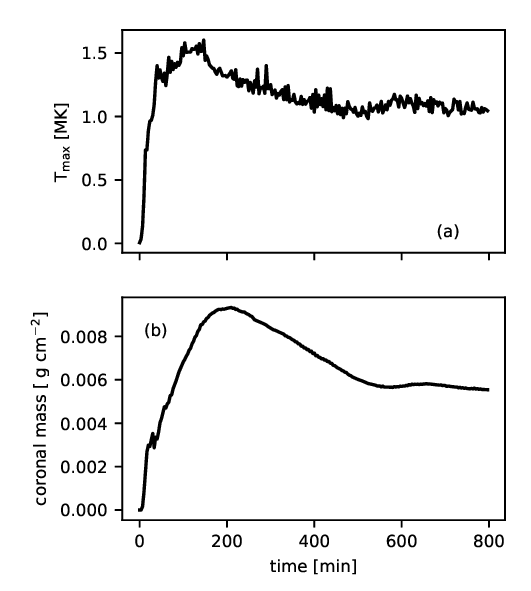}
    \caption{Temporal evolution of (a) the maximum temperature and (b) the coronal mass normalized by the cross sectional area at the photosphere.}
    \label{fig:time_evolution}
\end{figure}

The system underwent both the evaporation ($t<200$ min) and condensation ($200<t<500$ min) phases before attaining a quasi-steady state ($t>500$ min) (Fig. \ref{fig:time_evolution}). 
The coronal mass in Fig. \ref{fig:time_evolution}(b) was normalized by the cross-sectional area at the base, denoted as $A_0$, and defined as $\int \rho A/A_0 ds$, where integration was performed over the region where the temperature exceeded 10$^5$ K.
The particle flux during the evaporation and condensation phase was at most 5$\times$10$^{14}$ and -1$\times$10$^{14}$ cm$^{-2}$ s$^{-1}$, respectively. 
In the quasi-steady state, the model reproduced a warm loop with an apex temperature of approximately 1.1 MK, featuring sharp transition regions between the chromosphere and the corona (the minimum temperature scale height of $\sim$ 70 km on average).
The average electron density at the apex was 6.2 $\times$ 10$^8$ cm$^{-3}$, while the average length of the coronal loop was 79 Mm. 
These characteristics aligned with those of warm loops that have been extensively studied since the work of \cite{1999ApJ...515..842A}. 
At T=10$^5$ K, the velocity fluctuation ($\sqrt{\langle v_{\rm LOS}^2 \rangle} \sim \sqrt{\langle v_{\phi}^2 \rangle/2}$) measured approximately 25 km s$^{-1}$, consistent with the typical non-thermal line width observed in the transition region of the quiet sun \citep[e.g.][]{2001A&A...374.1108P}.
There are no noticeable trends in the vertical flow (i.e. $|\langle v_{\rm s} \rangle|<$ 1 km s$^{-1}$) within the transition region, although lower-temperature EUV lines typically exhibit a redshift \citep[e.g.][]{1999ApJ...522.1148P}.

\begin{figure}
	\includegraphics[width=9cm]{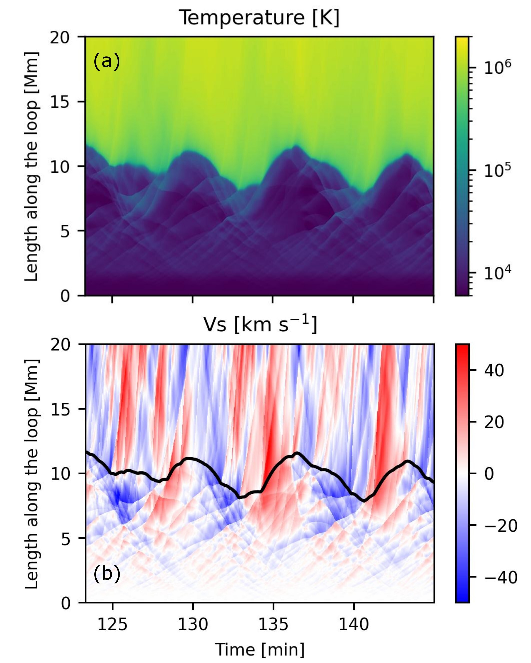}
    \caption{Time distance diagram of (a) temperature and (b) velocity along the field line.
    The black solid line in panel (b) indicates the contour at $T = 10^5$ K that represent the transition region.}
    \label{fig:td_diagram}
\end{figure}

Two significant dynamics pertaining to the transition region are noteworthy: the collision of shocks and evaporation/condensation processes.
Firstly, the transition region exhibited vertical motion in response to the interaction with MHD shocks emanating from the photosphere (see Fig. \ref{fig:td_diagram}). 
The typical altitude of the transition region measured approximately 9 Mm, with temporal variations of up to 3 Mm. 
These motions have previously been interpreted as spicule motion \citep{1982SoPh...75...35H, 1999ApJ...514..493K}.
This spicular dynamics did not change significantly through the whole calculation after the formation of the hot corona. 
Secondly, the shock-induced heating coincided with evaporation and condensation processes (see Fig. \ref{fig:time_evolution}b), resulting in the circulation of materials between the corona and the chromosphere. 
These phenomena contributed significantly to the dynamic ionization and recombination processes occurring within the region, rendering them pivotal factors for consideration in the examination of emission originating from the transition region. 
Subsequent subsections of this study will delve into these aspects in greater detail.

\subsection{Ionization fractions in the evaporation phase}

\begin{figure}
	\includegraphics[]{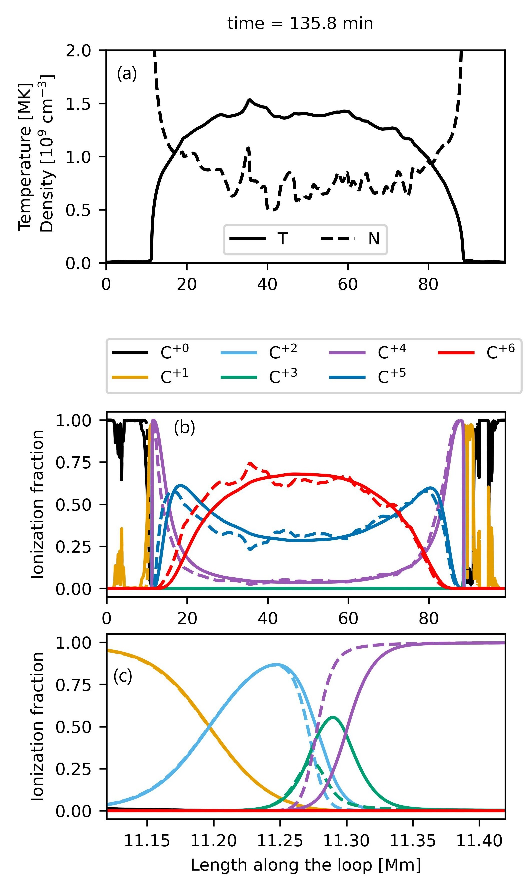}
    \caption{(a) coronal temperature (solid line) and density (dashed line) as a function of $s$ at t=135.8 min. (b) A snapshot of ionization fractions of C ions.
    (c) Zoomed-in view of (b) focused on the transition region.
    For panels (b) and (c), solid lines depict the NEI fractions, while dashed lines represent the ionization fraction in equilibrium state.
    }
    \label{fig:ion_frac_s}
\end{figure}

Significant departures from ionization equilibrium were observed in the middle of corona as well as near the transition region. 
Figure \ref{fig:ion_frac_s} presented a snapshot of the temperature, density, and ionization fractions of C ions during the evaporation phase. 
The spatial distribution of ionization fractions exhibited two notable features.
First, in the middle corona, the spatial profile of ionization fractions appeared considerably smoother than that in the equilibrium state (Fig. \ref{fig:ion_frac_s}b). 
This phenomenon can be attributed to the longer ionization time scale, which prevents the ionization fraction from promptly responding to temperature fluctuations induced by shocks and acoustic waves. 
The corresponding temperature and density fluctuations can be found in the figure \ref{fig:ion_frac_s}a.
Previous simulations with time-dependent heating rates have demonstrated that NEI effects can arise due to temperature fluctuations, even in the absence of mass motion \citep{1982ApJ...255..783M}. 
Observations have also indicated that the variability time scales are often constrained by ionization processes, regardless of the underlying atmospheric dynamics \citep{1989SoPh..122..245G}.
Second, the peak of the ionization fraction for Li- and Na-like ions (e.g., \ion{C}{4} in Fig. \ref{fig:ion_frac_s}c) was frequently displaced to higher altitudes. 
Consequently, the distributions of \ion{C}{3} and \ion{C}{5} were also shifted to higher altitudes.

\begin{figure}
	\includegraphics[width=\columnwidth]{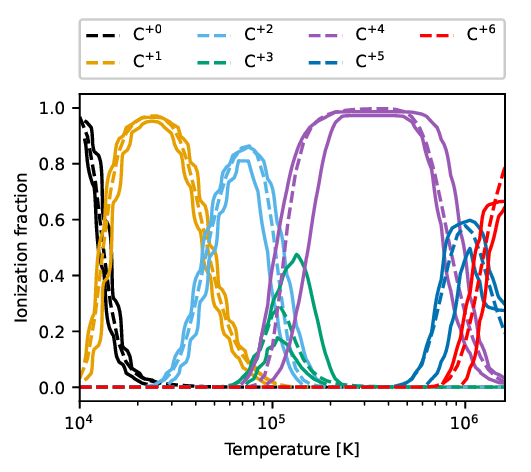}
    \caption{Ionization fractions of C ions as a function of temperature.
    Solid lines depict 1 sigma intervals of PDF of the NEI fractions.
    Dashed lines represent the fractions in ionization equilibrium.
    }
    \label{fig:ion_frac_te}
\end{figure}

The aforementioned deviations in the transition region were also evident in the probability distribution function (PDF) of the ionization fraction in temperature space (Fig. \ref{fig:ion_frac_te}).
Due to the effects of NEI, the ionization fraction did not solely depend on the local temperature, leading to a certain distribution with finite width at a specific temperature.
We discretized the temperature space into bins of $\Delta \log T = 0.05$ and computed PDFs, denoted as $F_{\rm i}$, for each temperature bin as follows:
\begin{equation}
F_{\rm i}(T;x) = P(T; n_{\rm i} \le x),
\end{equation}
where $i$ serves as the index for ion species;
$P(n_{\rm i} \le x)$ represents the probability that a particular ionization fraction, $n_{\rm i}$, is smaller than $x$ at temperature $T$.
To estimate PDF in the evaporation phase, we used the data in ionization phase (135 $\leq t \leq $ 145 min).
The solid lines in Fig. \ref{fig:ion_frac_te} depicted the ranges of 1-sigma levels for PDF ($F_{\rm i} \in [0.17,0.83]$), whereas the dashed lines illustrate the ionization fraction at the equilibrium state.
The over-fraction and shift toward higher temperature was found for \ion{C}{4} while other C ions almost followed the ionization fraction in equilibrium.
These characteristics differ from those documented in previous studies (\cite{2013ApJ...767...43O, 2015ApJ...802....5O, 2016ApJ...817...46M}), where substantial deviations in the ionization fraction were observed, even for ions among non Li- and Na-like ions. This discrepancy can be mainly attributed to the lower density in the transition region in their simulations, while the impact of numerical diffusion cannot be disregarded. The outcomes obtained from the coarse grid ($\sim$10 km width), demonstrating finer resolution than the earlier multi-dimensional simulations, revealed marked deviations from ionization equilibrium due to numerical diffusion, notably noticeable among non Li- and Na-like ions.

\begin{figure*}
	\includegraphics[width=\linewidth]{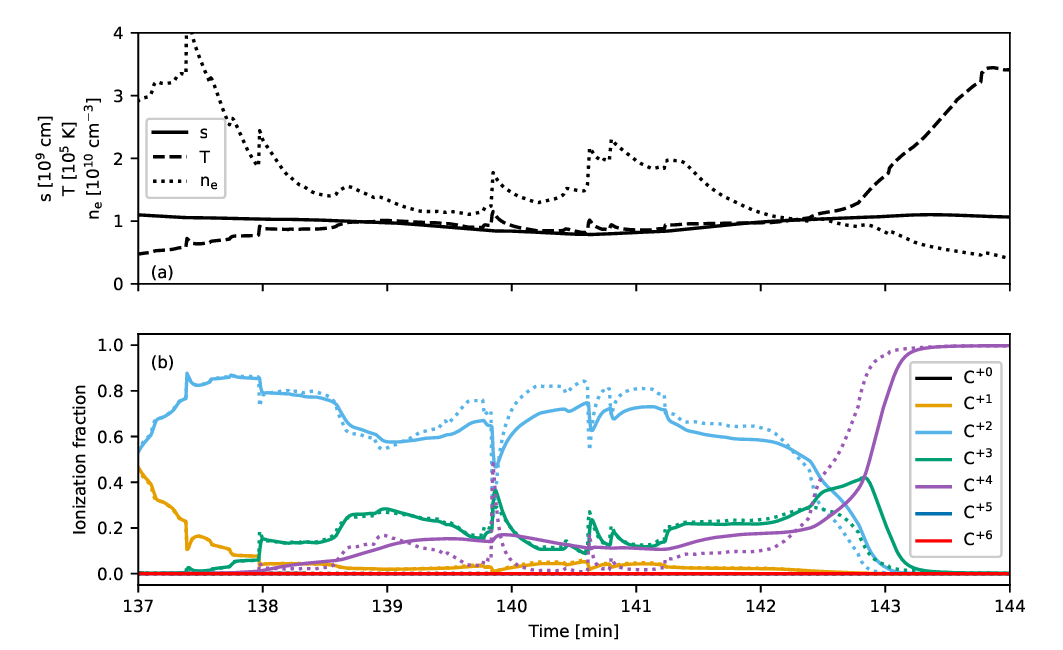}
    \caption{ (a) Temperature (dashed line), electron density (dotted line), position (solid line), and (b) ionization fractions of a traced particle in the evaporation phase.
    The solid and the dotted lines in panel (b) indicate the ionization fraction in non-equilibrium and equilibrium, respectively.
    }
    \label{fig:ion_frac_trace}
\end{figure*}

The reason for the enhancement of \ion{C}{4} fraction can be primarily attributed to the evaporation from the chromosphere to the corona.
In figure \ref{fig:ion_frac_trace}, we depicted evolution of a certain plasma parcel co-moving with fluid to trace its temperature, electron density, position, and ionization fractions.
From $t=$ 137.5 min, the ionization fraction of \ion{C}{4} increased as the plasma temperature increased to its formation temperature.
As time went on, the plasma parcel evaporated to coronal temperature from $t=$ 142.5 min in 40 sec, and then, the ionization fraction started to decrease.
Because the evaporation time scale of $\sim$ 40 sec was comparable to the recombination time scale, the ionization fraction stayed larger than that in equilibrium during the evaporation process.
The similar over-fractionation in Li- and Na-like ions can be found in the steady flow solution \citep{1989ApJ...338.1131N, 1990ApJ...362..370S, 2004ApJ...606.1239P}.
The over-fraction of \ion{C}{4} is also shown in the evaporation phase in nanoflare-heated loops \citep{1993ApJ...402..741H} and impulsive phase in flaring loop \citep{2004A&A...425..287B}.
Throughout the evaporation process, this plasma parcel encountered shocks at $t\sim$ 140 min and 140.8 min. Upon interaction with the shocks, the \ion{C}{4} ionization fraction increased spontaneously. Despite these collisions occurring within a 10-second timeframe, which is shorter than the ionization time scales, the \ion{C}{4} ionization fraction closely tracked the equilibrium fraction.
Considering that \ion{C}{3} and \ion{C}{5} displayed over- and under-fraction, respectively, the \ion{C}{4} ionization fraction was likely maintained by an enhancement of ionization and a reduction in recombination during the shock passage.
Nano-flare-generated waves entering to the transition region from the corona could also modify the ionization fraction \citep{1993ApJ...402..741H}, although we did not observe this phenomena probably because our model lacks nanoflares. 

\subsection{Ionization fractions in the condensation and quasi-steady phase}

\begin{figure*}
	\includegraphics[width=\linewidth]{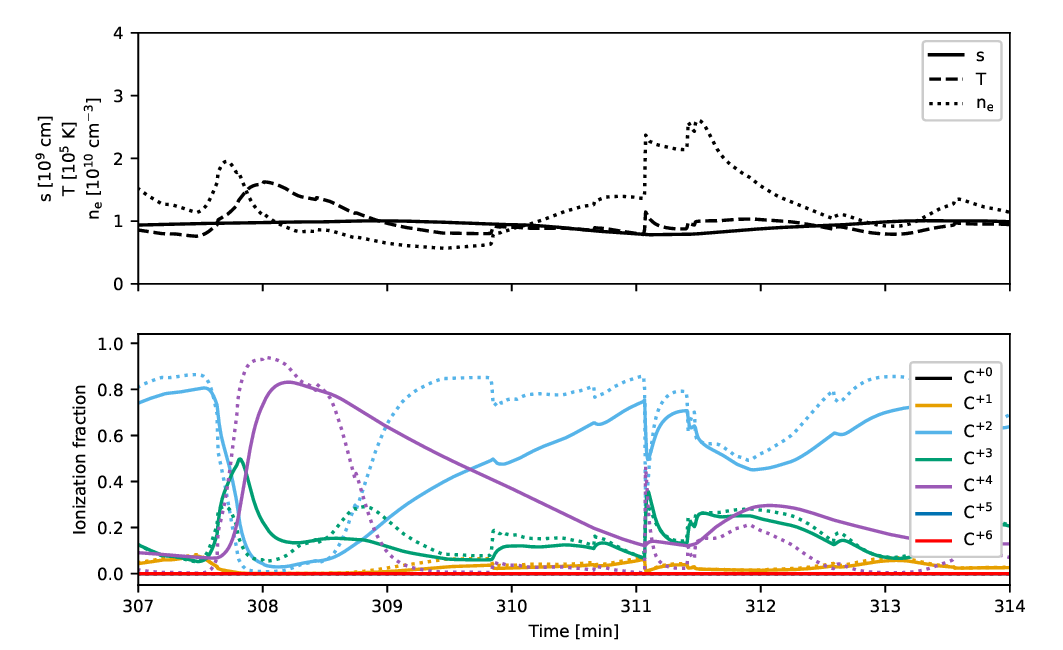}
    \caption{ The same format as in the figure \ref{fig:ion_frac_trace} in condensation phase.
    }
    \label{fig:ion_frac_trace_cond}
\end{figure*}

During the condensation phase, the \ion{C}{4} ions exhibited, on average, an under-fraction.
Figure \ref{fig:ion_frac_trace_cond} displayed the pertinent physical characteristics of the condensing plasma parcel. 
Due to the finite duration required for the recombination process from \ion{C}{5} to \ion{C}{4}, which exceeded the condensation timescale, the ionization fraction of \ion{C}{4} remained below that of the equilibrium state. 
This under-fraction is qualitatively consistent with the simulation by \cite{1993ApJ...402..741H} during the condensation phase in a nanoflare-heated loop and by \cite{1989ApJ...338.1131N, 1990ApJ...362..370S} in the down flowing siphon flow. 
Contrary, the over-fraction is reported in the inflowing model of \cite{2004ApJ...606.1239P}, which might be due to the difference in atmospheric model used, boundary conditions at the corona, or amplitude of mass flux.
It was observed that the plasma parcel experienced a collision with a shock at $t\sim$ 311 min, during which the \ion{C}{4} ionization fraction once again closely followed the equilibrium fraction, as was noted during the evaporation phase. Despite the plasma undergoing repeated evaporation (e.g., at $t\sim$ 307.8 min) and condensation episodes during the whole condensation phase, the average ionization fraction of \ion{C}{4} was consistently smaller than that in the equilibrium state.

The shocks in themselves do not alter the average ionization fraction of \ion{C}{4} when averaging over time.
In the quasi-steady state, where there is no tendency towards continuous evaporation or condensation, we can delineate the impacts of temperature and density fluctuations caused by shocks on NEI. Since there are no substantial deviations from ionization equilibrium in the quasi-steady state, we assert that the processes of evaporation/condensation are necessary to induce over/under fractions within the present model.

\subsection{Synthetic UV intensity}

The synthetic UV intensities obtained using the coronal approximation were generally smaller than those derived with NEI by a maximum of 40\% during the evaporation phase (Table \ref{tab:ratio}). The UV intensity computation under the coronal approximation is governed by the following formula:
\begin{eqnarray}
    I(\lambda) = \int Ab(X) C(T, \lambda, N_{\rm e}) N_{\rm e} N_{\rm H} ds. \label{eq:intensity}
\end{eqnarray}
Here, $Ab(X)$ represents the abundance of element $X$, assuming photospheric abundance. The function $C(T, \lambda, N_{\rm e})$ corresponds to the contribution functions for each spectral line, calculated using the CHIANTI software.
For NEI plasma intensities, we modified the integrand in eq. (\ref{eq:intensity}) by multiplying it with the ionization fraction ratio, $N_{\rm i;NEI}/N_{\rm i;EI}$. Among several spectral lines, we specifically highlighted lines with a ratio $I_{\rm EI}/I_{\rm NEI}$ less than 0.8 during the evaporation phase and an intensity greater than 10 erg cm$^{-2}$ s$^{-1}$ sr$^{-1}$.
While the effect of NEI significantly enhanced the intensity through the over-fraction during the evaporation phase, our model still exhibited discrepancies between $R$ (defined as $I_{\rm EI}/I_{\rm NEI}$) and $I_{\rm th}/I_{\rm obs}$, as observed in quiet sun \citep{1995ApJ...455L..85J, 2023MNRAS.521.4696D}. 
The observational data utilized to determine the intensity ratio is obtained from a complex combination of several instruments documented in \cite{1995ApJ...455L..85J} and \cite{2023MNRAS.521.4696D}. In short, these datasets were obtained from the quiet region. However, it is noteworthy that the data employed in \cite{1995ApJ...455L..85J} are spatially unresolved, potentially allowing contributions from active regions to be present.
The under-fraction observed during the condensation and quasi-steady phases resulted in the ratio $R<1.1$ at most.
The observed intensity ratios are still significantly smaller than those from our model, although we can not directly compare the observations in quiet regions and the model in active region loop.
It is anticipated that the influence of non-equilibrium ionization (NEI) will be more pronounced in the regions characterized by lower density, particularly in the quiet region.

\begin{table*}
\centering
\caption{Comparison of intensities from ionization fractions in equilibrium and non-equilibrium during the evaporation phase.}
\label{tab:ratio}
\begin{tabular}{lrrrrrr}
\hline
Ion (Seq)  & $\lambda$ [\AA] & $I_{\rm EI}$ & $I_{\rm NEI}$ & $R_{\rm e}$&$R_{\rm c}$& $I_{\rm th}/I_{\rm obs}$\\
\hline
\ion{C}{4} (Li)&1548.19&3452.0&4328.2&0.8&1.1&0.197$^{\rm a}$, 0.31-1.28$^{\rm b}$\\
\ion{C}{4} (Li)&1550.77&1725.0&2162.2&0.8&1.1&0.34-1.01$^{\rm b}$\\
\ion{C}{4} (Li)&384.17&27.0&43.0&0.6&1.0&\\
\ion{C}{4} (Li)&312.42&19.4&30.8&0.6&1.0&\\
\ion{C}{4} (Li)&419.71&25.1&39.2&0.6&1.1&\\
\ion{C}{4} (Li)&384.03&15.0&23.9&0.6&1.0&\\
\ion{N}{5} (Li)&1238.82&417.2&599.7&0.7&1.1&0.198$^{\rm a}$, 0.28$^{\rm b}$\\
\ion{N}{5} (Li)&1242.80&208.6&299.7&0.7&1.1&0.17-0.40$^{\rm b}$\\
\ion{N}{5} (Li)&247.71&10.5&15.6&0.7&1.0&\\
\ion{N}{5} (Li)&209.27&7.2&10.7&0.7&&\\
\ion{N}{5} (Li)&266.38&7.5&11.4&0.7&&\\
\ion{O}{6} (Li)&1031.91&3586.7&4675.5&0.8&1.0&0.524$^{\rm a}$,0.5-0.71$^{\rm b}$\\
\ion{O}{6} (Li)&1037.61&1785.8&2326.5&0.8&1.0&0.41-0.76$^{\rm b}$\\
\ion{O}{6} (Li)&173.08&143.9&185.6&0.8&1.0&\\
\ion{O}{6} (Li)&184.12&95.9&125.5&0.8&1.0&\\
\ion{O}{6} (Li)&172.94&80.1&103.4&0.8&1.0&\\
\ion{O}{6} (Li)&183.94&48.4&63.3&0.8&1.0&\\
\ion{Si}{4} (Na)&1128.34&33.5&41.7&0.8&1.0&0.16-0.19$^{\rm b}$\\
\ion{Si}{4} (Na)&1122.48&18.8&23.4&0.8&1.0&\\
\ion{S}{6} (Na)&933.38&120.4&163.6&0.7&1.0&0.31-0.59$^{\rm b}$\\
\ion{S}{6} (Na)&944.52&60.0&81.5&0.7&1.0&0.58-1.15$^{\rm b}$\\
\ion{S}{6} (Na)&712.67&8.2&11.2&0.7&&0.4$^{\rm b}$\\
\hline
\multicolumn{7}{p{10cm}}{
Note. Ion - emitting ion; Seq - ion isoelectronic sequence; $\lambda$ - wavelength; $I_{\rm EI}$ and $I_{\rm NEI}$ - predicted intensity in EI and NEI during the evaporation phase [ergs cm$^{-2}$ s$^{-1}$ sr$^{-1}$]; $R_{\rm e},R_{\rm c}$ - ratio of $I_{\rm EI}$ to $I_{\rm NEI}$ during the evaporation and condensation phase, respectively; $I_{\rm th}/I_{\rm obs}$ - ratio of theoretical predictions of intensity to observed intensity ($^{\rm a}$ \cite{1995ApJ...455L..85J} and $^{\rm b}$ \cite{2023MNRAS.521.4696D}).}\\
\end{tabular}
\end{table*}


\subsection{The effect of broadening technique of the transition region}

In our study, we implemented a numerical broadening technique for the transition region to mitigate numerical diffusion effects near the lower transition region, which was identified as a factor reducing the impacts NEI. 
To investigate the impact of this technique, we conducted simulations with a broader transition region, specifically setting $T_{\rm c}=0.2$ MK. 
The results of this simulation revealed a increase in the ratio of $R$, such as an increase from 0.6 to 0.8 for \ion{C}{4} ions 384.17 \AA, and similar increases were observed for other Li- and Na-like ions. 
This increase in the intensity ratio can be attributed to the broadening of the lower transition region, which results in a longer dynamical time scale for traversing the layer compared to realistic conditions, thus potentially reducing the influence of NEI.
Consequently, we consider the intensity ratio obtained under our parameter settings to represent an upper limit in this context.

\subsection{Mass circulation}

The magnitude of over-fraction evolved through the mass circulation process occurring between the corona and the chromosphere, as our model has unveiled distinct ionization fractions across various phases.
Our model disclosed that evaporation occurred prior to $t<$ 200 min, followed by the onset of condensation until approximately $t\sim$ 500 min. 
The duration of these mass cycles could be affected by the choice of abundance thorough the radiative cooling function. Because we used the photospheric abundance in this study, the cooling time scale could be shorter when using the coronal abundance instead.
NEI effect on the radiative cooling function may slightly increase the cooling time scale of the coronal loop \citep{2003A&A...401..699B} if we include the feedback effects.
Subsequently, the system maintained a quasi-steady state, at least until $t=$ 800 min, though it is worth noting that the continuity of this quasi-steady state beyond that point is not guaranteed.
Similar models with significantly extended computational duration have elucidated the cyclic evolution of coronal loops \citep{2019ApJ...885..164W}, a phenomenon that could be attributed to the thermal instability inherent in coronal loops \citep{1982A&A...108L...1K}.
While providing a quantitative estimate for the amplitude of over-fraction remains challenging, it is reasonable to anticipate that the ratio $R$ presented during the evaporation phase (as detailed in Table \ref{tab:ratio}) would undergo increase when taking the condensation phase into consideration.

\section{Conclusions}
In this study we have conducted 1.5-dimensional MHD simulations for Alfv\'{e}n-wave-heated coronal loops simultaneously solving a series of NEI equations for several ion species. 
After introducing the Alfv\'{e}nic fluctuations at the foot point of the loop, the system experiences the evaporation and condensation phases before it reaches the quasi-steady state.
During the evaporation phase, the over-fractionation of Li- and Na-like ions results in the intensity ratios, $R$, as low as 0.6. Conversely, during the condensation and quasi-steady phases, the under-fractionation leads to $R$ values as high as 1.1. These pronounced fluctuations in ionization fractions are primarily attributed to the processes of evaporation and condensation between the corona and the chromosphere. Interestingly, the collision with shocks do not significantly deviate the \ion{C}{4} ion fraction from the ionization equilibrium.

While our model has successfully demonstrated a reduction in the intensity ratios $R$ during the evaporation phase, a noticeable gap between the model and observational data still persists. 
This may suggest that the heating mechanism in quiet regions exhibits more sporadic behavior, leading to a more dynamic mass circulation between the chromosphere and the corona. In such instances, deviations from equilibrium ionization could be pronounced, resulting in a reduced intensity ratio during the evaporation phase. Concurrently, this leads to an increased intensity ratio during the condensation phase. Therefore, it is not straightforward to assert that stronger mass circulation unequivocally leads to a smaller intensity ratio when averaged over the entire period.
Bridging this gap also necessitates further investigations into atomic physics aspects, such as density effects, photoionization, and charge transfer, which collectively have the potential to align the theoretically predicted UV intensities of transition region spectral lines with observations \citep{2023MNRAS.521.4696D}.

The observed over-fractionation of Li- and Na-like ions holds significant scientific interest, as it may serve as an indicator of mass motions closely linked to coronal heating mechanisms and mass loss processes. This phenomenon gains relevance in the context of the wealth of supporting evidence for impulsive heating events that drive the cycle of evaporation and condensation in the solar atmosphere \citep{2006SoPh..234...41K}. Consequently, the extent of over-fractionation could potentially offer valuable constraints on the amplitude and frequency of such impulsive heating events \citep{2011ApJS..194...26B}.
Furthermore, it's worth noting that the solar wind are known to expand the UV-observed solar atmosphere via the effects of NEI \citep{1964spre.conf..730N,1978ApJ...226..698M,1979ApJ...229L.101D}. Consequently, the degree of over-fraction could be considered as a metric for mass loss processes in the Sun. Importantly, anomalies in the ionization states of Li- and Na-like ions have also been identified in stellar atmospheres \citep{2002A&A...385..968D}, suggesting that the observation of these ions may provide valuable insights into the heating and mass-loss mechanisms in other stars.

Multi-dimensional simulations that incorporate NEI effects are crucial for conducting a quantitative comparison between theoretical models and complex observational data \citep{2013ApJ...767...43O, 2015ApJ...802....5O, 2016ApJ...817...46M}. However, such simulations remain a formidable challenge, even with the current computational resources at our disposal. This challenge arises from the necessity to accurately resolve the narrow transition region when solving the advection equations for Li- and Na-like ions.
The width of the transition region where those ions form, determined by the Field length or $\sqrt{\kappa T/|\rho {\cal L}|}$ \citep{1990ApJ...358..375B}, was sometimes going down to $\sim$ 20 km at $T\sim10^5$ K in our simulation. Notably, this width is broadened by a factor of $(T_c/10^5~{\rm K})^{5/2} \sim 2.8$ with our broadening technique, effectively reducing the width to approximately 7 km in the realistic situation. To achieve a high-resolution representation of this thin layer, typically requiring at least 10 grid points to mitigate numerical diffusion, we find it necessary to employ the adaptive mesh refinement schemes, even in one-dimensional simulations \citep{2003A&A...401..699B}.
The TRAC method proposed in \cite{2019ApJ...873L..22J} is a robust technique for accurately determining coronal density regardless of grid size. However, it may not improve the situation when solving the advection equations in ionization fractions because the transition region is resolved with only a few grids in this method. Insufficient resolution to capture the transition region leads to pronounced numerical diffusion, giving rise to unphysical deviations from ionization equilibrium. 

The extent of over-fractionation is likely contingent on the mass flux or the rate of mass exchange between the corona and the chromosphere. However, our current model does not comprehensively explore this particular parameter. To conduct a thorough investigation of these parameters, it would be advantageous to employ time-steady solutions \citep{1989ApJ...338.1131N,1990ApJ...362..370S,2004ApJ...606.1239P, 2020ApJ...901..150G}.
By deriving the over- and under-fractionation behaviors of Li- and Na-like ions as functions of mass flux, we could potentially establish valuable constraints on coronal heating mechanisms and mass loss rates. These constraints would be particularly informative when compared with UV observations in future observations from Solar Orbiter/SPICE \citep{2013SPIE.8862E..0FF} or Solar-C/EUVST \citep{2020SPIE11444E..0NS}, which cover large wavelength range.

\begin{acknowledgments}
We extend our heartfelt gratitude to the reviewer for their insightful comments and constructive feedback, which greatly enhanced the quality of this manuscript.
This work was supported by JSPS KAKENHI Grant Number JP23K03456.
This study was carried by using the computational resource of the Center for Integrated Data Science, Institute for Space-Earth Environmental Research, Nagoya University through the joint research program.
CHIANTI is a collaborative project involving George Mason University, the University of Michigan (USA), University of Cambridge (UK) and NASA Goddard Space Flight Center (USA).
\end{acknowledgments}

%



\software{CHIANTI \citep{2021ApJ...909...38D}
          }



\appendix

\begin{figure}
	\includegraphics[width=9cm]{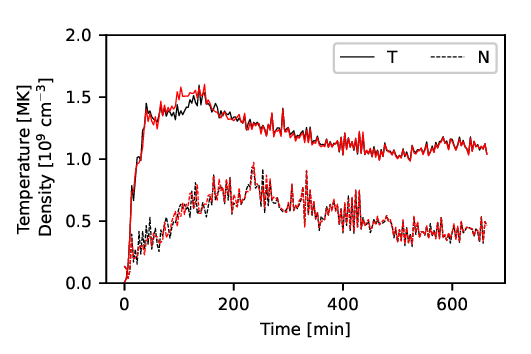}
    \caption{Resolution dependence on the evolution of the maximum temperature (solid) and coronal density at apex (dashed). The red and black lines represent the results from coarse simulation runs with 10 and 5 km, respectively. }
    \label{fig:resolution_coarse}
\end{figure}

We performed two distinct convergence tests to explore the resolution dependency of our findings.
The first test pertains to the coarse simulation, which served as the initial conditions for the fine-grid simulations.
In this coarse simulation, we employed a grid size of 5 km near the transition region, whereas the original grid size was 10 km.
Figure \ref{fig:resolution_coarse} illustrates the entire evolution of the maximum temperature (solid black line) and the density at the apex (dashed black line), alongside the results obtained with a 10 km grid size (depicted by red lines).
Based on this comparison, we infer that the maximum temperature and the coronal density exhibit minimal sensitivity to resolutions finer than 10 km.

\begin{figure}
	\includegraphics[width=9cm]{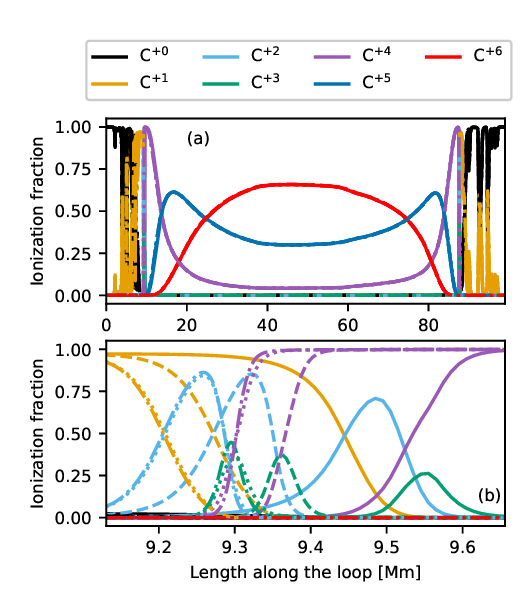}
    \caption{Resolution dependence on the ionization fraction of Carbon ions at t=145 min. The solid, dashed, dotted, and dash-dotted lines represent the results with 10, 5, 2.5, and 1.25 km, respectively. The different colors indicate the level of ionization. The panel (b) is a zoomed-in view of (a) focused on the transition region. }
    \label{fig:resolution_on_ionizfrac}
\end{figure}

The second test pertains to the fine-scale simulations employed for computing the ionization fractions.
We varied the resolution from 10 to 1.25 km and depicted the ionization fraction at t=145 min ($\sim$ 20 min after the starting time of fine-scale simulation) in Figure \ref{fig:resolution_on_ionizfrac}.
The colors represent the ionization fraction of Carbon ions, following the same manner as depicted in Figure \ref{fig:ion_frac_s}.
Results obtained with grid sizes of 10, 5, 2.5, and 1.25 km are denoted by solid, dashed, dotted, and dash-dotted lines, respectively.
A comparison between the results obtained with grid sizes of 2.5 km and 1.25 km indicates that the ionization fraction has nearly converged with a grid size of 1.25 km.
While the ionization fraction in the coronal region (\ion{C}{5}, \ion{C}{6}, \ion{C}{7}) exhibits convergence with a grid size of 10 km, a resolution of 1.25 km is necessary to achieve convergence for transition region ions such as \ion{C}{3} and \ion{C}{4}.

\bibliography{myref}{}

\hyphenation{Post-Script Sprin-ger}
\begin{thebibliography}{}
\expandafter\ifx\csname natexlab\endcsname\relax\def\natexlab#1{#1}\fi
\providecommand{\url}[1]{\href{#1}{#1}}
\providecommand{\dodoi}[1]{doi:~\href{http://doi.org/#1}{\nolinkurl{#1}}}
\providecommand{\doeprint}[1]{\href{http://ascl.net/#1}{\nolinkurl{http://ascl.net/#1}}}
\providecommand{\doarXiv}[1]{\href{https://arxiv.org/abs/#1}{\nolinkurl{https://arxiv.org/abs/#1}}}

\bibitem[{{Antolin} \& {Shibata}(2010)}]{2010ApJ...712..494A}
{Antolin}, P., \& {Shibata}, K. 2010, ApJ, 712, 494, \dodoi{10.1088/0004-637X/712/1/494}

\bibitem[{{Antolin} {et~al.}(2008){Antolin}, {Shibata}, {Kudoh}, {Shiota}, \& {Brooks}}]{2008ApJ...688..669A}
{Antolin}, P., {Shibata}, K., {Kudoh}, T., {Shiota}, D., \& {Brooks}, D. 2008, \apj, 688, 669, \dodoi{10.1086/591998}

\bibitem[{{Aschwanden} {et~al.}(1999){Aschwanden}, {Newmark}, {Delaboudini{\`e}re}, {Neupert}, {Klimchuk}, {Gary}, {Portier-Fozzani}, \& {Zucker}}]{1999ApJ...515..842A}
{Aschwanden}, M.~J., {Newmark}, J.~S., {Delaboudini{\`e}re}, J.-P., {et~al.} 1999, \apj, 515, 842, \dodoi{10.1086/307036}

\bibitem[{{Balsara} {et~al.}(2009){Balsara}, {Rumpf}, {Dumbser}, \& {Munz}}]{2009JCoPh.228.2480B}
{Balsara}, D.~S., {Rumpf}, T., {Dumbser}, M., \& {Munz}, C.-D. 2009, Journal of Computational Physics, 228, 2480, \dodoi{10.1016/j.jcp.2008.12.003}

\bibitem[{{Begelman} \& {McKee}(1990)}]{1990ApJ...358..375B}
{Begelman}, M.~C., \& {McKee}, C.~F. 1990, \apj, 358, 375, \dodoi{10.1086/168994}

\bibitem[{{Bradshaw} {et~al.}(2004){Bradshaw}, {Del Zanna}, \& {Mason}}]{2004A&A...425..287B}
{Bradshaw}, S.~J., {Del Zanna}, G., \& {Mason}, H.~E. 2004, \aap, 425, 287, \dodoi{10.1051/0004-6361:20040521}

\bibitem[{{Bradshaw} \& {Klimchuk}(2011)}]{2011ApJS..194...26B}
{Bradshaw}, S.~J., \& {Klimchuk}, J.~A. 2011, \apjs, 194, 26, \dodoi{10.1088/0067-0049/194/2/26}

\bibitem[{{Bradshaw} \& {Mason}(2003)}]{2003A&A...401..699B}
{Bradshaw}, S.~J., \& {Mason}, H.~E. 2003, \aap, 401, 699, \dodoi{10.1051/0004-6361:20030089}

\bibitem[{{Burton} {et~al.}(1971){Burton}, {Jordan}, {Ridgeley}, \& {Wilson}}]{1971RSPTA.270...81B}
{Burton}, W.~M., {Jordan}, C., {Ridgeley}, A., \& {Wilson}, R. 1971, Philosophical Transactions of the Royal Society of London Series A, 270, 81, \dodoi{10.1098/rsta.1971.0063}

\bibitem[{{Chitta} {et~al.}(2012){Chitta}, {van Ballegooijen}, {Rouppe van der Voort}, {DeLuca}, \& {Kariyappa}}]{2012ApJ...752...48C}
{Chitta}, L.~P., {van Ballegooijen}, A.~A., {Rouppe van der Voort}, L., {DeLuca}, E.~E., \& {Kariyappa}, R. 2012, \apj, 752, 48, \dodoi{10.1088/0004-637X/752/1/48}

\bibitem[{{Del Zanna} {et~al.}(2021){Del Zanna}, {Dere}, {Young}, \& {Landi}}]{2021ApJ...909...38D}
{Del Zanna}, G., {Dere}, K.~P., {Young}, P.~R., \& {Landi}, E. 2021, \apj, 909, 38, \dodoi{10.3847/1538-4357/abd8ce}

\bibitem[{{Del Zanna} {et~al.}(2002){Del Zanna}, {Landini}, \& {Mason}}]{2002A&A...385..968D}
{Del Zanna}, G., {Landini}, M., \& {Mason}, H.~E. 2002, \aap, 385, 968, \dodoi{10.1051/0004-6361:20020164}

\bibitem[{{Del Zanna} \& {Mason}(2018)}]{2018LRSP...15....5D}
{Del Zanna}, G., \& {Mason}, H.~E. 2018, Living Reviews in Solar Physics, 15, 5, \dodoi{10.1007/s41116-018-0015-3}

\bibitem[{{Dud{\'\i}k} {et~al.}(2014){Dud{\'\i}k}, {Del Zanna}, {Dzif{\v{c}}{\'a}kov{\'a}}, {Mason}, \& {Golub}}]{2014ApJ...780L..12D}
{Dud{\'\i}k}, J., {Del Zanna}, G., {Dzif{\v{c}}{\'a}kov{\'a}}, E., {Mason}, H.~E., \& {Golub}, L. 2014, \apjl, 780, L12, \dodoi{10.1088/2041-8205/780/1/L12}

\bibitem[{{Dud{\'\i}k} {et~al.}(2017){Dud{\'\i}k}, {Dzif{\v{c}}{\'a}kov{\'a}}, {Meyer-Vernet}, {Del Zanna}, {Young}, {Giunta}, {Sylwester}, {Sylwester}, {Oka}, {Mason}, {Vocks}, {Matteini}, {Krucker}, {Williams}, \& {Mackovjak}}]{2017SoPh..292..100D}
{Dud{\'\i}k}, J., {Dzif{\v{c}}{\'a}kov{\'a}}, E., {Meyer-Vernet}, N., {et~al.} 2017, \solphys, 292, 100, \dodoi{10.1007/s11207-017-1125-0}

\bibitem[{{Dufresne} {et~al.}(2023){Dufresne}, {Del Zanna}, \& {Mason}}]{2023MNRAS.521.4696D}
{Dufresne}, R.~P., {Del Zanna}, G., \& {Mason}, H.~E. 2023, \mnras, 521, 4696, \dodoi{10.1093/mnras/stad794}

\bibitem[{{Dupree}(1972)}]{1972ApJ...178..527D}
{Dupree}, A.~K. 1972, \apj, 178, 527, \dodoi{10.1086/151813}

\bibitem[{{Dupree} {et~al.}(1979){Dupree}, {Moore}, \& {Shapiro}}]{1979ApJ...229L.101D}
{Dupree}, A.~K., {Moore}, R.~T., \& {Shapiro}, P.~R. 1979, \apjl, 229, L101, \dodoi{10.1086/182938}

\bibitem[{{Einfeldt} {et~al.}(1991){Einfeldt}, {Roe}, {Munz}, \& {Sjogreen}}]{1991JCoPh..92..273E}
{Einfeldt}, B., {Roe}, P.~L., {Munz}, C.~D., \& {Sjogreen}, B. 1991, Journal of Computational Physics, 92, 273, \dodoi{10.1016/0021-9991(91)90211-3}

\bibitem[{{Fludra} {et~al.}(2013){Fludra}, {Griffin}, {Caldwell}, {Eccleston}, {Cornaby}, {Drummond}, {Grainger}, {Greenway}, {Grundy}, {Howe}, {McQuirk}, {Middleton}, {Poyntz-Wright}, {Richards}, {Rogers}, {Sawyer}, {Shaughnessy}, {Sidher}, {Tosh}, {Beardsley}, {Burton}, {Marshall}, {Waltham}, {Woodward}, {Appourchaux}, {Philippon}, {Auchere}, {Buchlin}, {Gabriel}, {Vial}, {Sch{\"u}hle}, {Curdt}, {Innes}, {Meining}, {Peter}, {Solanki}, {Teriaca}, {Gyo}, {B{\"u}chel}, {Haberreiter}, {Pfiffner}, {Schmutz}, {Carlsson}, {Haugan}, {Davila}, {Jordan}, {Thompson}, {Hassler}, {Walls}, {Deforest}, {Hanley}, {Johnson}, {Phelan}, {Blecha}, {Cottard}, {Paciotti}, {Autissier}, {Allemand}, {Relecom}, {Munro}, {Butler}, {Klein}, \& {Gottwald}}]{2013SPIE.8862E..0FF}
{Fludra}, A., {Griffin}, D., {Caldwell}, M., {et~al.} 2013, in Society of Photo-Optical Instrumentation Engineers (SPIE) Conference Series, Vol. 8862, Solar Physics and Space Weather Instrumentation V, ed. S.~{Fineschi} \& J.~{Fennelly}, 88620F, \dodoi{10.1117/12.2027581}

\bibitem[{{Gilly} \& {Cranmer}(2020)}]{2020ApJ...901..150G}
{Gilly}, C.~R., \& {Cranmer}, S.~R. 2020, \apj, 901, 150, \dodoi{10.3847/1538-4357/abb1ad}

\bibitem[{{Golub} {et~al.}(1989){Golub}, {Hartquist}, \& {Quillen}}]{1989SoPh..122..245G}
{Golub}, L., {Hartquist}, T.~W., \& {Quillen}, A.~C. 1989, \solphys, 122, 245, \dodoi{10.1007/BF00912995}

\bibitem[{{Hansteen}(1993)}]{1993ApJ...402..741H}
{Hansteen}, V. 1993, \apj, 402, 741, \dodoi{10.1086/172174}

\bibitem[{{Hollweg} {et~al.}(1982){Hollweg}, {Jackson}, \& {Galloway}}]{1982SoPh...75...35H}
{Hollweg}, J.~V., {Jackson}, S., \& {Galloway}, D. 1982, \solphys, 75, 35, \dodoi{10.1007/BF00153458}

\bibitem[{{Johnston} \& {Bradshaw}(2019)}]{2019ApJ...873L..22J}
{Johnston}, C.~D., \& {Bradshaw}, S.~J. 2019, \apjl, 873, L22, \dodoi{10.3847/2041-8213/ab0c1f}

\bibitem[{{Judge} {et~al.}(1995){Judge}, {Woods}, {Brekke}, \& {Rottman}}]{1995ApJ...455L..85J}
{Judge}, P.~G., {Woods}, T.~N., {Brekke}, P., \& {Rottman}, G.~J. 1995, \apjl, 455, L85, \dodoi{10.1086/309815}

\bibitem[{{Klimchuk}(2006)}]{2006SoPh..234...41K}
{Klimchuk}, J.~A. 2006, \solphys, 234, 41, \dodoi{10.1007/s11207-006-0055-z}

\bibitem[{{Kudoh} \& {Shibata}(1999)}]{1999ApJ...514..493K}
{Kudoh}, T., \& {Shibata}, K. 1999, ApJ, 514, 493, \dodoi{10.1086/306930}

\bibitem[{{Kuin} \& {Martens}(1982)}]{1982A&A...108L...1K}
{Kuin}, N.~P.~M., \& {Martens}, P.~C.~H. 1982, \aap, 108, L1

\bibitem[{{Lionello} {et~al.}(2009){Lionello}, {Linker}, \& {Miki{\'c}}}]{2009ApJ...690..902L}
{Lionello}, R., {Linker}, J.~A., \& {Miki{\'c}}, Z. 2009, \apj, 690, 902, \dodoi{10.1088/0004-637X/690/1/902}

\bibitem[{{MacNeice} {et~al.}(1984){MacNeice}, {McWhirter}, {Spicer}, \& {Burgess}}]{1984SoPh...90..357M}
{MacNeice}, P., {McWhirter}, R.~W.~P., {Spicer}, D.~S., \& {Burgess}, A. 1984, \solphys, 90, 357, \dodoi{10.1007/BF00173963}

\bibitem[{{Mariska} {et~al.}(1982){Mariska}, {Doschek}, {Boris}, {Oran}, \& {Young}}]{1982ApJ...255..783M}
{Mariska}, J.~T., {Doschek}, G.~A., {Boris}, J.~P., {Oran}, E.~S., \& {Young}, T.~R., J. 1982, \apj, 255, 783, \dodoi{10.1086/159877}

\bibitem[{{Mariska} {et~al.}(1978){Mariska}, {Feldman}, \& {Doschek}}]{1978ApJ...226..698M}
{Mariska}, J.~T., {Feldman}, U., \& {Doschek}, G.~A. 1978, \apj, 226, 698, \dodoi{10.1086/156652}

\bibitem[{{Mart{\'\i}nez-Sykora} {et~al.}(2016){Mart{\'\i}nez-Sykora}, {De Pontieu}, {Hansteen}, \& {Gudiksen}}]{2016ApJ...817...46M}
{Mart{\'\i}nez-Sykora}, J., {De Pontieu}, B., {Hansteen}, V.~H., \& {Gudiksen}, B. 2016, \apj, 817, 46, \dodoi{10.3847/0004-637X/817/1/46}

\bibitem[{{Matsumoto} \& {Kitai}(2010)}]{2010ApJ...716L..19M}
{Matsumoto}, T., \& {Kitai}, R. 2010, ApJ, 716, L19, \dodoi{10.1088/2041-8205/716/1/L19}

\bibitem[{{Matsumoto} \& {Shibata}(2010)}]{2010ApJ...710.1857M}
{Matsumoto}, T., \& {Shibata}, K. 2010, ApJ, 710, 1857, \dodoi{10.1088/0004-637X/710/2/1857}

\bibitem[{{Matsumoto} \& {Suzuki}(2014)}]{2014MNRAS.440..971M}
{Matsumoto}, T., \& {Suzuki}, T.~K. 2014, \mnras, 440, 971, \dodoi{10.1093/mnras/stu310}

\bibitem[{{Meyer} {et~al.}(2012){Meyer}, {Balsara}, \& {Aslam}}]{2012MNRAS.422.2102M}
{Meyer}, C.~D., {Balsara}, D.~S., \& {Aslam}, T.~D. 2012, \mnras, 422, 2102, \dodoi{10.1111/j.1365-2966.2012.20744.x}

\bibitem[{{Moriyasu} {et~al.}(2004){Moriyasu}, {Kudoh}, {Yokoyama}, \& {Shibata}}]{2004ApJ...601L.107M}
{Moriyasu}, S., {Kudoh}, T., {Yokoyama}, T., \& {Shibata}, K. 2004, \apjl, 601, L107, \dodoi{10.1086/381779}

\bibitem[{{Moriyasu} \& {Shibata}(2004)}]{2004ESASP.575...80M}
{Moriyasu}, S., \& {Shibata}, K. 2004, in ESA Special Publication, Vol. 575, SOHO 15 Coronal Heating, ed. R.~W. {Walsh}, J.~{Ireland}, D.~{Danesy}, \& B.~{Fleck}, 80

\bibitem[{{Neupert}(1964)}]{1964spre.conf..730N}
{Neupert}, W.~M. 1964, in Space Research Conference, 730

\bibitem[{{Noci} {et~al.}(1989){Noci}, {Spadaro}, {Zappala}, \& {Antiochos}}]{1989ApJ...338.1131N}
{Noci}, G., {Spadaro}, D., {Zappala}, R.~A., \& {Antiochos}, S.~K. 1989, \apj, 338, 1131, \dodoi{10.1086/167263}

\bibitem[{{Oba} {et~al.}(2017){Oba}, {Riethm{\"u}ller}, {Solanki}, {Iida}, {Quintero Noda}, \& {Shimizu}}]{2017ApJ...849....7O}
{Oba}, T., {Riethm{\"u}ller}, T.~L., {Solanki}, S.~K., {et~al.} 2017, \apj, 849, 7, \dodoi{10.3847/1538-4357/aa8e44}

\bibitem[{{Olluri} {et~al.}(2013){Olluri}, {Gudiksen}, \& {Hansteen}}]{2013ApJ...767...43O}
{Olluri}, K., {Gudiksen}, B.~V., \& {Hansteen}, V.~H. 2013, \apj, 767, 43, \dodoi{10.1088/0004-637X/767/1/43}

\bibitem[{{Olluri} {et~al.}(2015){Olluri}, {Gudiksen}, {Hansteen}, \& {De Pontieu}}]{2015ApJ...802....5O}
{Olluri}, K., {Gudiksen}, B.~V., {Hansteen}, V.~H., \& {De Pontieu}, B. 2015, \apj, 802, 5, \dodoi{10.1088/0004-637X/802/1/5}

\bibitem[{{Peter}(2001)}]{2001A&A...374.1108P}
{Peter}, H. 2001, \aap, 374, 1108, \dodoi{10.1051/0004-6361:20010697}

\bibitem[{{Peter} \& {Judge}(1999)}]{1999ApJ...522.1148P}
{Peter}, H., \& {Judge}, P.~G. 1999, \apj, 522, 1148, \dodoi{10.1086/307672}

\bibitem[{{Pietarila} \& {Judge}(2004)}]{2004ApJ...606.1239P}
{Pietarila}, A., \& {Judge}, P.~G. 2004, \apj, 606, 1239, \dodoi{10.1086/383176}

\bibitem[{{Raymond} \& {Doyle}(1981)}]{1981ApJ...245.1141R}
{Raymond}, J.~C., \& {Doyle}, J.~G. 1981, \apj, 245, 1141, \dodoi{10.1086/158889}

\bibitem[{{Shimizu} {et~al.}(2020){Shimizu}, {Imada}, {Kawate}, {Suematsu}, {Hara}, {Tsuzuki}, {Katsukawa}, {Kubo}, {Ishikawa}, {Watanabe}, {Toriumi}, {Ichimoto}, {Nagata}, {Hasegawa}, {Yokoyama}, {Watanabe}, {Tsuno}, {Korendyke}, {Warren}, {De Pontieu}, {Boerner}, {Solanki}, {Teriaca}, {Schuehle}, {Matthews}, {Long}, {Thomas}, {Hancock}, {Reid}, {Fludra}, {Auch{\`e}re}, {Andretta}, {Naletto}, {Poletto}, \& {Harra}}]{2020SPIE11444E..0NS}
{Shimizu}, T., {Imada}, S., {Kawate}, T., {et~al.} 2020, in Society of Photo-Optical Instrumentation Engineers (SPIE) Conference Series, Vol. 11444, Space Telescopes and Instrumentation 2020: Ultraviolet to Gamma Ray, ed. J.-W.~A. {den Herder}, S.~{Nikzad}, \& K.~{Nakazawa}, 114440N, \dodoi{10.1117/12.2560887}

\bibitem[{{Shoda} \& {Takasao}(2021)}]{2021A&A...656A.111S}
{Shoda}, M., \& {Takasao}, S. 2021, \aap, 656, A111, \dodoi{10.1051/0004-6361/202141563}

\bibitem[{{Spadaro} {et~al.}(1990){Spadaro}, {Zappala}, {Antiochos}, {Lanzafame}, \& {Noci}}]{1990ApJ...362..370S}
{Spadaro}, D., {Zappala}, R.~A., {Antiochos}, S.~K., {Lanzafame}, G., \& {Noci}, G. 1990, \apj, 362, 370, \dodoi{10.1086/169273}

\bibitem[{{van Ballegooijen} {et~al.}(1998){van Ballegooijen}, {Nisenson}, {Noyes}, {L{\"o}fdahl}, {Stein}, {Nordlund}, \& {Krishnakumar}}]{1998ApJ...509..435V}
{van Ballegooijen}, A.~A., {Nisenson}, P., {Noyes}, R.~W., {et~al.} 1998, \apj, 509, 435, \dodoi{10.1086/306471}

\bibitem[{{Vernazza} \& {Raymond}(1979)}]{1979ApJ...228L..89V}
{Vernazza}, J.~E., \& {Raymond}, J.~C. 1979, \apjl, 228, L89, \dodoi{10.1086/182910}

\bibitem[{{Washinoue} \& {Suzuki}(2019)}]{2019ApJ...885..164W}
{Washinoue}, H., \& {Suzuki}, T.~K. 2019, \apj, 885, 164, \dodoi{10.3847/1538-4357/ab48ec}

\end{thebibliography}
\bibliographystyle{aasjournal}



\end{document}